\documentstyle[preprint,aps,psfig]{revtex}

\newlength{\dinwidth}
\newlength{\dinmargin}
\newcommand{\be}{\begin{eqnarray}}
\newcommand{\ee}{\end{eqnarray}}

\begin{document}

\draft
\preprint{\begin{tabular}{r}
  {\bf hep-ph/0006322}\\
  YUMS 00-05
\end{tabular}}
\vspace{1cm}

\title{
Hadronic Invariant Mass Spectrum \\
in $ B \rightarrow X_u l \nu $ Decay
with Lepton Energy Cut}
\author{
K. K. Jeong${~^a}$, C. S. Kim${~^{a,b}}$\footnote{kim@kimcs.yonsei.ac.kr,~~~
http://phya.yonsei.ac.kr/\~{}cskim/}
and Yeong Gyun Kim${~^c}$
}
\vspace{0.5cm}
\address{
$a:$ Department of Physics, Yonsei University and IPAP, Seoul 120--749, Korea\\ 
$b:$ Department of Physics, University of Wisconsin, Madison, WI 53706, USA\\ 
$c:$ YITP, Kyoto University, Kyoto 606--8502, Japan 
}

\vspace{0.5cm}
\date{\today}
\maketitle
\begin{abstract}

\noindent
We discuss the implications of charged lepton energy cut to
the hadronic invariant mass spectrum in charmless semileptonic $B$ decays.
Charged-lepton energy cut is inevitable in order to remove
secondary leptonic events such as $b \rightarrow c,\tau \rightarrow l$,
and to identify the chaged leptons at detectors experimentally.
We consider three possible lepton energy cuts, $E_l^{\rm cuts} = 0.6,1.5,2.3$
GeV, and found that
with the most probable cuts $E_l^{\rm cut} = 1.5$ GeV and $M_X^{\rm max} =
1.5~(1.86)$ GeV, $45 \sim 60 \% ~(58 \sim 67 \%)$ of decay events survive.
Therefore, $B \to X_u l \nu$ decay events can be efficiently distinguished
from $B \to X_c l \nu$ decay events.
We also discuss the possible model dependence on the results. 
\end{abstract}

\pacs{ }

\narrowtext

\section{Introduction}

The determination of the CKM parameter $|V_{ub}|$ is important
for constructing the so-called unitary triangle. 
It is hard to determine $|V_{ub}|$ from semileptonic $B$-meson
decays because Cabbibo dominated decay mode $B \to X_c l \nu$ 
obscures the $B \to X_u l \nu$ mode. The traditional method for
extracting $|V_{ub}|$ from experimental data involves a study 
of the charged lepton energy spectrum in inclusive
semileptonic $B$ decays, $B \rightarrow X_{u,c} l \nu$ \cite{cleo}.
The $b \rightarrow u$ events are selected above charm threshold,
$i.e.$ for lepton energies $E_l$ above 
$(M^2_B-M^2_D)/(2M_B) \approx$ 2.3 \,GeV.
However, this cut on $E_l$ is not very efficient (only less than 10$\%$ of
$b \rightarrow u$ events survive) and also the dependence of the lepton
energy spectrum on perturbative and non-perturbative QCD corrections is
the strongest in this end-point region
\cite{cabibbo,chay,manohar,neubert,bigi}.

As an alternative, the determination of $|V_{ub}|$ 
may come from the measurement of the
hadronic invariant mass spectrum \cite{barger} in the region $M_X < M_D$.
For $B \rightarrow X_{c} l \nu$ decays, one necessarily has 
$M_X > M_D = 1.86$ GeV. Therefore, if we impose a condition $M_X < M_D$,
the resulting events come only from $B \rightarrow X_u l \nu$ decay and 
most of the $B \rightarrow X_u l \nu$ decays are expected to lie in this 
region. There is an experimental problem to be expected, though --
the charm-leaking of misidentified charmed particles below the kinematic
$b \rightarrow c$ threshold. To avoid this leakage, one may concentrate 
on hadronic invariant mass below a certain value $M_X^{\rm max}(< M_D)$, say
$M_X^{\rm max} = 1.5$ GeV. The detailed studies of this method were
performed in Refs. \cite{dike,af,fn}. The integrated fraction of events was
introduced,
\begin{eqnarray}
\Phi(M_X^{\rm max})=\frac{1}{\Gamma(B \rightarrow X_u l \nu)}
   \int^{M_X^{\rm max}}_{0} \mbox{d}M_X \frac{\mbox{d}\Gamma}{\mbox{d}M_X}~,
\end{eqnarray}
and studied in its sensitivity to the three basic parameters, $\mu^2_\pi$,
$m_b$ and $\alpha_s$.

However, we cannot apply the above results to the experimental data directly.
From the practical point of view, the leptons with low energy, $i.e.$ less
than  0.6 GeV, cannot be experimentally identified within the dectectors. And
also a larger lepton energy cut might be needed to 
select `prompt leptonic events' $(b \rightarrow l)$ from 
`secondary leptonic events' 
$(b \rightarrow c \rightarrow l,~ \tau \rightarrow l)$.
An experimental method such as 
the technique of neutrino reconstruction \cite{cleo2},
which would be used to measure hadronic invariant mass 
directly and inclusively, may require 
a lower cut on the charged lepton energy.
Therefore, the hadronic invariant mass spectrum 
would be affected by the various lepton energy cuts.
In this Letter, we study the effects of lepton energy cut on the hadronic
invariant mass spectrum in inclusive charmless semileptonic $B$ decays and 
discuss their implications.

\section{Differential decay rate}

At the parton level, the most general hadronic tensor 
for $B \rightarrow X_u l \nu$ 
is given by
\be
W_{\mu \nu} (p,v) &=& W_1 (v \cdot p, p^2) 
(p_\mu v_\nu+p_\nu v_\mu-g_{\mu \nu} v \cdot p
-i \epsilon_{\mu \nu \alpha \beta} p^\alpha v^\beta) \nonumber \\
&-& W_2 (v \cdot p, p^2) g_{\mu \nu} + W_3 (v \cdot p, p^2) v_\mu v_\nu
\nonumber \\
&+& W_4 (v \cdot p, p^2) (p_\mu v_\nu + p_\nu v_\mu)
+ W_5 (v \cdot p, p^2) p_\mu p_\nu~.
\ee
Here $v$ is the $b$-quark velocity and $p$ is the total parton momentum.
The total momentum carried by the leptons is $q=m_b v-p$. 
At the tree level, $W_1 = 2\delta(p^2)$ and all other $W_{i\ne 1} = 0$.
We follow the $O(\alpha_s)$ corrections to the hadronic tensor
from the paper of De Fazio and Neubert \cite{fn}.
Introducing the scaling variables
\be
x \equiv {2 E_l \over m_b}~,~~~\hat{p}^2 \equiv {p^2 \over m_b^2}~~~{\rm and}~~~
z \equiv {2 v \cdot p \over m_b}~,
\ee
where $E_l$ is the charged-lepton energy which is defined in the
$B$-meson rest frame, 
the triple differential decay rate is given by
\be
\frac{\mbox{d}^3 \Gamma}{\mbox{d}x\,\mbox{d}z\,\mbox{d}\hat{p}^2} = 
12 \Gamma_0 \bigg\{ &(1&+\bar{x}-z)(z-\bar{x}-\hat{p}^2){m_b^2 \over 2} W_1
+(1-z+\hat{p}^2){m_b \over 2} W_2 \nonumber \\
&+& (\bar{x}(z-\bar{x})-\hat{p}^2){m_b \over 4} 
(W_3 + 2 m_b W_4 + m_b^2 W_5) \bigg\},
\ee
where $\bar{x} \equiv 1-x$, and
\be
\Gamma_0 = {G_F^2 |V_{ub}|^2 m_b^5 \over 192 \pi^3}.
\ee

In terms of the parton variables, the total invariant mass of the
hadronic final state is given by
\be
s_{_H} = p^2 + 2 \bar{\Lambda} v \cdot p + \bar{\Lambda}^2,
\ee
where $\bar{\Lambda} \equiv M_B - m_b$.
Using  relation (6), and the scaling variables,
$$\hat{s}_{_H} \equiv s_{_H} / m_b^2
~~~{\rm and}~~~\epsilon \equiv \bar{\Lambda} / m_b ~;$$ 
we find the following double differential decay rate
\be
\frac{\mbox{d}^2 \Gamma}{\mbox{d}x\,\mbox{d}\hat{s}_{_H}} = 
\int \mbox{d}z {\mbox{d}^3 \Gamma \over \mbox{d}x\,\mbox{d}z\,d\hat{p}^2}
\bigg|_{\displaystyle \hat{p}^2 =\hat{s}_{_H} -\epsilon z-\epsilon^2} ~,
\ee
where the phase space for the relevant variables is given by
\be
{\hat s_{_H}-\epsilon^2+\bar x^2 \over \epsilon + \bar x} \le &z& \, \le
{\hat s_{_H}-\epsilon^2 \over \epsilon}, \nonumber \\ 
{\epsilon (1+\epsilon)-\hat s_{_H} \over \epsilon} \le &x& \, \le 1, \\
\mbox{for}~~~~~~~~~~~~~~~~~~\epsilon^2 \le &\hat s_{_H}& \, 
\le \epsilon (1+\epsilon)~; \nonumber \\
\nonumber \\
{\hat s_{_H}-\epsilon^2+\bar x^2 \over \epsilon + \bar x} \le &z& \, \le
1-\epsilon+{\hat s_{_H} \over 1+\epsilon},\nonumber \\ 
0 \le &x& \, \le {(1+\epsilon)^2-\hat s_{_H} \over 1+\epsilon}, \\
\mbox{for}~~~~~~~~~~\epsilon (1+\epsilon)
\le &\hat s_{_H}& \, \le (1+\epsilon)^2. \nonumber 
\ee
If we integrate the above double differential decay rate over the variable $x$,
we get the single differential decay rate for $\hat s_{_H}$ 
\be
{\mbox{d} \Gamma  \over \mbox{d} \hat s_{_H}} = 
\int \mbox{d} x\, \frac{\mbox{d}^2 \Gamma}{\mbox{d}x\,\mbox{d}\hat{s}_{_H}}\,\,
,  \ee
which reproduces numerically the result of Ref. \cite{fn}.

In order to obtain the physical decay distributions, we should also consider 
the non-perturbative corrections. 
The physical decay distributions are obtained from convolution of
the above parton level spectra
with non-perturbative shape function $F(k_+)$,
which governs the light-cone momentum distribution of 
the heavy quark inside the $B$-meson \cite{neubert,bigi}. 
The convolution of parton spectra with this function is such that
for the decay distributions
the $b$-quark mass $m_b$ is replaced by the momentum dependent
mass, $m_b + k_+$, and similarly the parameter $\bar \Lambda = M_B - m_b$
is replaced by $\bar \Lambda - k_+$.
Here $k_+$ can take values between $-m_b$ and $\bar \Lambda$, with
a distribution centered around $k_+=0$ and with a characteristic width of
${\cal O}(\Lambda)$.
Then, the scaling variables $x$, $\hat s_{_H}$ and $\epsilon$ 
are replaced by the new variables
\be
x_q \equiv {2 E_l \over M_B - q_+}~,~~~~ 
\hat s_{{_H},q} \equiv {s_{_H} \over (M_B-q_+)^2}~~~~{\rm and}~~~
\epsilon_q \equiv {q_+ \over M_B - q_+}~,
\ee
where $q_+ \equiv \bar \Lambda - k_+$.
The physical double differential decay rate for 
the charged-lepton energy and the hadronic invariant mass is given by
\be
{\mbox{d}^2 \Gamma \over \mbox{d} E_l \, \mbox{d} s_{_H}} =
2 \int_0^{q_+^{\rm max}} \mbox{d} q_+ 
{F(\bar \Lambda - q_+) \over (M_B-q_+)^3}~
{\mbox{d}\Gamma \over 
\mbox{d}x \,\mbox{d}\hat s_{_H}} (x_q,\hat s_{{_H},q},\epsilon_q),
\ee
where $q_+^{\rm max} = \mbox{min}(M_B-2 E_l,\sqrt{s_{_H}})$.
Finally, the physical distribution for the hadronic invariant mass can be
obtained in two different ways;
\be
{\mbox{d} \Gamma \over \mbox{d} s_{_H}} &=& 
\int \mbox{d} E_l \,
{\mbox{d}^2 \Gamma \over \mbox{d} E_l \, \mbox{d} s_{_H}}
\\ \nonumber \\
&=&
\int_0^{\sqrt{s_{_H}}} \mbox{d} q_+ 
{F(\bar \Lambda - q_+) \over (M_B-q_+)^2}~
{\mbox{d}\Gamma \over 
\mbox{d}\hat s_{_H}} (\hat s_{{_H},q},\epsilon_q),
\ee
which should give the same results. We note that,
after the implementation of such Fermi motion, now the kinematic variables take
values in the entire phase space determined by hadron kinematics, $i.e.$
$0 \le E_l \le  M_B/2$ and $0 \le s_{_H} ~(\equiv M_X^2) \le M_B^2$.
Eqs. (12-14) are our starting point for the numerical calculations.

\section{Numerical Analyses and Conclusions}

To perform the numerical calculation we should choose a specific form
for the shape function $F(k_+ \equiv \bar \Lambda - q_+)$. 
It is subject to the constraints on the moments of the function
$$
A_n = \langle k_+^n \rangle \equiv
\int \mbox{d} k_+ k_+^n F(k_+) ,
$$
which are given by the expectation values of
local heavy quark operators.
In practice we know only the size of
the first few moments; one finds
\be
A_0=1,~~ A_1=0~~~{\rm and}~~~A_2={1 \over 3} \mu^2_\pi~,
\ee
where $\mu_\pi^2$ is the average momentum squared of 
the $b$ quark inside the $B$-meson.
The parameters $\bar \Lambda$ and $\mu^2_\pi$ were obtained by the HQET and
QCD sum rules in Refs. \cite{ball,neubert1};
\be
\bar \Lambda &=& 0.4 \sim 0.6,
\nonumber \\
\mu_{\pi}^2 = 0.6 \pm 0.1~~\cite{ball}~~~~&{\rm or}&~~~~
\mu_{\pi}^2 = 0.10 \pm 0.05~~\cite{neubert1}.
\ee
One then chooses some reasonable ansatz for $F(k_+)$; its parameters
are adjusted so as to reproduce its known moments.
Several functional forms for this function have been suggested 
in the literature \cite{bigi,lee,kim,kagan}.
We adopt the simple form of Ref. \cite{kagan},
\be
F(k_+) = N (1-x)^a e^{(1+a)x}~~~{\rm with}~~~ 
x={k_+ \over \bar \Lambda} \le 1,
\ee
which is such that $A_1 = 0$ by construction
(neglecting exponentially small terms in $m_b / \Lambda$),
whereas the condition for $A_0$ fixes the normalization $N$.
The parameter $a$ can be related to the second moment, yielding
$A_2 = {\bar \Lambda}^2 / (1+a)$. A typical choice of values is 
$m_b = 4.8$\,GeV and $a = 1.29$, corresponding to
$\bar \Lambda \approx 0.48$\,GeV and $\mu_\pi^2 \approx 0.3$\,GeV$^2$.
Below, we keep the parameter `$a$' fixed, and consider the three choices 
$m_b = 4.65,\,4.8$ and 4.95 GeV. These choices correspond to
the following sets, 
\be
(\bar \Lambda({\rm GeV}), \mu_{\pi}^2({\rm GeV}^2)) =
(0.63,0.52),~~ (0.48,0.30)~~~{\rm and}~~~ (0.33,0.14),
\ee
respectively. We fix the QCD coupling $\alpha_s = 0.22$.

Now the calculation of the hadronic invariant mass spectrum
is straightforward. 
The 3-dimensional plot of double differential decay rate,
\be
\frac{\mbox{d}\Gamma}{\mbox{d}M_X dE_l} = 
2M_X \frac{\mbox{d}\Gamma}{\mbox{d}s_{_H} \mbox{d}E_l} 
\ee
is shown in Fig. 1, with $m_b = 4.8$ GeV.
The lines in the bottom plane show the contour lines.
The three straight lines in contour figure, which are the projections of
three curved lines in the 3-dimensional figure, 
indicate the charged-lepton energy cuts,
$i.e.~$ $E_l^{\rm cut} = 0.6, 1.5, 2.3$ GeV, respectively.
From the figure, one can find that 
a significant fraction ($\sim 90\%$)
of the decay events lie below the charmed states, 
$i.e.$ have $M_X < 1.86$ GeV.
This is in sharp contrast with the usual cut on charged-lepton energy $E_l$.
Due to measurement errors, there will be a tail from $b \rightarrow c$
decays below $M_D$. 
To avoid this leakage, one can concentrate on hadronic invariant
masses below a certain value $M_X^{\rm max} < M_D$.
With a lower cutoff on $M_X$ (say $M_X^{\rm max} = 1.5$ GeV  or 1.6 GeV),
the majority of decays still appears below the $M_X^{\rm max}$ region.

But in reality we cannot apply the above results to experimental data
directly. 
In $B \to Xl\nu$ semileptonic decay experiment, the secondary electrons
such as $b \to c \to l$ cascade decay contaminates
the signal of the pure $B \to X_c l\nu$ decay.
The secondary electrons have typically lower energy than the primary electrons
because  the mass of $D$-meson is much less than the mass of $B$-meson.
Therefore, only the electrons with energy greater than 1.5 GeV can be
considered to be purely from $B \to X_{c,u} l\nu$.
Using the double lepton tagging method, one can reconstruct the primary leptons
with energy less than 1.5 GeV \cite{doubletag}.
But in this method, both $B$-mesons from $\Upsilon(4S)$ should decay
semileptonically. So it gives much less data samples.
Also the experimental method such as neutrino reconstruction
technique, which would be used to measure the hadronic invariant mass
inclusively, requires a lower cut on the charged lepton energy.
Therefore, we apply the charged-lepton energy cuts $E_l^{\rm cut} = 0.6, 1.5,
2.3$ GeV  on the hadronic invariant mass spectrum.
The cut $E_l^{\rm cut} = 0.6$ GeV corresponds to the double lepton tagging
data, and $E_l^{\rm cut} = 1.5$ GeV corresponds to the normal single
lepton tagging data. We also apply $E_l^{\rm cut} = 2.3$ GeV to compare the
data with the previously measured charged lepton energy data
as the kinematic boundary to separate the signal from $b \to cl\nu$ decay.

Fig. 2 shows the hadronic invariant mass 
distribution with varying $m_b$ mass as 4.65, 4.8,
4.95~GeV. It is normalized in units of ($|V_{ub}|^2 \times$ GeV).
Note that in the $b \to u$ semileptonic decay,
the leptons have quite large energies so that $M_X$ 
distribution with $E_l^{\rm cut} = 0.6$~GeV is not
much different from the distribution without any lepton energy cut.
In the case of $E_l^{\rm cut} = 2.3$~GeV, we can see that there
is no decay event above  $M_X \ge 1.86$~GeV. This is consistent with the
kinematics that leptons with energy greater than 2.3~GeV are purely from $b
\to ul\nu$ decay. And also we can see that this lepton energy cut is very
inefficient because only a small part of the decays survive this cut,
as mentioned earlier.

Fig. 3 shows the integrated decay rate up to 
$M_X^{\rm max}$, 
\be
\Gamma(M_X^{\rm max}) = \int_0^{M_X^{\rm max}} 
\mbox{d}M_X \frac{\mbox{d}\Gamma}{\mbox{d}M_X} .
\ee
The numerical values are summarized in Tables I, II and III.
Here we can see that if we consider total decay rate with 
$E_l^{\rm cut}$ = 2.3 GeV,
only $9 \sim 14 \%$ of decay events remain. 
When we use $M_X^{\rm max} = M_D$ cut, 
$82 \sim 94 \%$ of decay events remain without any lepton energy cut,
and  $79 \sim 92 \% ~(58 \sim 67 \%)$ of decay events survive for
$E_l^{\rm cut} = 0.6~ (1.5)$ GeV.
In this case, there might be a difficulty to
draw the reliable conclusion because of the possible charm-leaking.
However, even in this case, the hard lepton energy cut may be helpful to 
reduce the contamination from the charm-leaking because
the lepton energy of $B \to X_c l \nu$ decays is softer compared to that of
$B \to X_u l \nu$ decays.
If we further reduce the maximum value of $M_X, ~i.e. ~M_X^{\rm max} =
1.5$~GeV, $60 \sim 81 \%$ of the events remain without lepton energy
cut, and $59 \sim 80 \%~ (45\% \sim 60\%$) of the events survive 
for $E_l^{\rm cut} = 0.6~(1.5)$~GeV.
As we mentioned previously, in the case of $E_l^{\rm cut}=0.6$~GeV,
both $B$-mesons should decay semileptonically to be identified as 
prompt leptonic events.
Since $E_l^{\rm cut}=2.3$~GeV gives only $\sim 10\%$ of the total decay rate,
the cut $E_l^{\rm cut}=1.5$~GeV would be our best option.

Now we discuss the dependence of the results
on the various input parameters
and the choice of the universal distribution function.
The dependence on  $\mu_\pi^2$ is found not so significant, compared to
the $\bar \Lambda$ (or equivalently $m_b$) dependence. 
The decay rates with the
parameters $m_b = 4.8$ GeV and $\mu_\pi^2 = 0.6$ GeV$^2$ are shown in Table IV.
By comparing Table IV with Tables I, II and III, 
we can see that the main uncertainty
in decay rates comes from the uncertainty in $\bar \Lambda$ (or $m_b$).
The dependence of the result on the $\alpha_s$ variation could be quite large
since the perturbative correction to the 
total decay rate is linear with the parameter $\alpha_s$,
and the size of $\alpha_s$ correction is almost 20\% of the leading 
approximation. From  Tables I, II, III and V, 
we can see that the effects of the variation of 
$\alpha_s$ from 0.22 to 0.35, with fixed $m_b = 4.8$~ GeV, are similar to
the case of the variation of $m_b$ from 4.8 to 4.65 GeV,
with fixed $\alpha_s = 0.22$.

Next, in order to estimate 
the possible dependence of the results on the choice of universal
distribution funtion, we adopt ACCMM \cite{accmm} model-induced distribution
function  ~\cite{bigi,kim}, 
\begin{equation}
F_{\rm ACCMM} (q_+) = \frac{1}{\sqrt{\pi}} \frac{1}{p_{_F}}
\exp\left\{ -\frac{1}{4} \left[ \frac{m_{sp}^2}{p_{_F} \; q_+} 
-\frac{q_+}{p_{_F}} \right]^2 \right\} .
\end{equation}
As seen, this shape function is dependent on the two parameters,
$p_{_F}$ and $m_{sp}$. We choose the three sets, 
$$
(p_{_F}, m_{sp}) =
(0.504, 0.251),~~~(0.383, 0.193)~~~
{\rm  and}~~~ (0.262, 0.136)~~~({\rm in~units~of~GeV}),
$$
which respectively correspond to the above three choices of $m_b =$ 4.65, 4.8
and 4.95 GeV with fixed `$a$' of Eq. (17).
The integrated decay rates $\Gamma(M_X^{\rm max})$ are summarized in
Tables VI and VII. 
The resulting decay rates are almost the same as the corresponding
results of Tables I and III. 
The differences in the decay rate for
two universal distribution functions are within 2\% only. 

Finally we note that the results would be affected by some resonance effects
around $M_X^{\rm max}$.
The real result would be a sum of all the exclusive decays in which 
a few resonances dominate at some specific $M_X$'s,
so the actual $M_X$ distribution will be with humps and
bumps, while our results are smoothed inclusive results using the duality.
However, our integrated results would be quite correct, once there is no
significant resonance around the region of $M_X$ cut. 
If $M_X^{\rm max}$ is around 1 GeV, 
then there are a few important resonances, but if $M_X$ is large enough, $e.g.$
1.5 GeV,  then there is no significant resonance from $b \rightarrow u$,
but rather there will be  decays with many pions.

In summary, we investigated the effects of charged lepton energy cut 
on the hadronic invariant mass spectrum for $B \rightarrow X_u l \nu$ decays
and their implications. As is well known, the
charged-lepton energy cut is experimentally inevitable in order to remove
secondary leptonic events such as $b \rightarrow c \rightarrow l,\tau
\rightarrow l$, and to identify the chaged leptons at detectors.
We found that with $E_l^{\rm cut} = 1.5$ GeV and $M_X^{\rm max} = 1.5~(1.86)$
GeV, $45 \sim 60 \% ~(58 \sim 67 \%)$ of decay events survive,
and with $E_l^{\rm cut} = 0.6$ GeV and $M_X^{\rm max} = 1.5~(1.86)$ GeV,
$59 \sim 80 \% ~(79 \sim 92 \%)$  survive.
Finally without $E_l^{\rm cut}$,
$60 \sim 81 \% ~(82 \sim 94 \%)$ of decay events survive with
$M_X^{\rm max} = 1.5~(1.86)$ GeV.
Therefore,
$B \to X_u l \nu$ decay events can be efficiently distinguished from
$B \to X_c l \nu$ decay events by using the hadronic recoil mass  and
the charged lepton energy  together.

\vspace{1.0cm}
\acknowledgements

We thank G. Cvetic for careful reading of the manuscript and his
valuable comments.
The work of C.S.K. was supported 
in part by the BSRI Program of MOE, Project No.
99-015-DI0032, 
and in part by the KRF Sughak-research program, 
Project No. 1997-011-D00015.
The work of K.K.J. was supported by Brain Korea 21 Project and by 
the KOSEF through Grant No. 1999-2-111-002-5.
The work of Y.G.K is supported by JSPS.

\newpage

\begin{table}[tb]
\caption{\label{haha} $\Gamma_{\rm total}$ with charged-lepton energy cuts 
with fixed $\alpha_s = 0.22$ and $a=1.29$. 
The values of $m_b= 4.65,~4.8,~4.95$ GeV correspond to the set of 
values of Eq. (18) for  $\bar \Lambda,~\mu_{\pi}^2$. } 
\[
\begin{array}{cccc}
\hline
m_b \mbox{(GeV)} & E_l^{\rm cut} \mbox{(GeV)} & \Gamma_{\rm total}
         (|V_{ub}|^2 \mbox{ GeV}) & \mbox{percentage} \\
\hline
4.8  & 0   & 0.503 & 100\% \\
     & 0.6 & 0.489 & 97\% \\
     & 1.5 & 0.336 & 67\% \\
     & 2.3 & 0.054 & 11\% \\
\hline
4.65 & 0   & 0.442 & 100\% \\
     & 0.6 & 0.429 & 97\% \\
     & 1.5 & 0.287 & 65\% \\
     & 2.3 & 0.039 &  9\% \\
\hline
4.95 & 0   & 0.574 & 100\% \\
     & 0.6 & 0.560 & 98\% \\
     & 1.5 & 0.395 & 69\% \\
     & 2.3 & 0.079 & 14\% \\
\hline
\end{array}
\]
\end{table}

\begin{table}[tb]
\caption{\label{haha1}
$\Gamma(M_X^{\rm max}=1.86 \mbox{GeV})$ with charged-lepton energy cuts.  }
\[
\begin{array}{cccc} 
\hline
m_b \mbox{(GeV)} & E_l^{\rm cut} \mbox{(GeV)} & \Gamma(M_X^{\rm max}=1.86
        \mbox{ GeV}) (|V_{ub}|^2 \mbox{ GeV}) & \mbox{percentage} \\
\hline
4.8  & 0   & 0.442 & 88\% \\
     & 0.6 & 0.432& 86\% \\
     & 1.5 & 0.312 & 62\% \\
     & 2.3 & 0.054 & 11\% \\
\hline
4.65 & 0   & 0.359 & 82\% \\
     & 0.6 & 0.351& 79\% \\
     & 1.5 & 0.254 & 58\% \\
     & 2.3 & 0.039 &  9\% \\
\hline
4.95 & 0   & 0.538 & 94\% \\
     & 0.6 & 0.526& 92\% \\
     & 1.5 & 0.381 & 67\% \\
     & 2.3 & 0.079 & 14\% \\
\hline
\end{array}
\]
\end{table}

\begin{table}[tb]
\caption{\label{haha2}
$\Gamma(M_X^{\rm max}=1.50 \mbox{GeV})$ with charged-lepton energy cuts. }
\[
\begin{array}{cccc} 
\hline
m_b \mbox{(GeV)} & E_l^{\rm cut} \mbox{(GeV)} & \Gamma(M_X^{\rm max}=1.50
        \mbox{ GeV}) (|V_{ub}|^2 \mbox{ GeV}) & \mbox{percentage} \\
\hline
4.8  & 0   & 0.348 & 69\% \\
     & 0.6 & 0.342 & 67\% \\
     & 1.5 & 0.261 & 52\% \\
\hline
4.65 & 0   & 0.263 & 60\% \\
     & 0.6 & 0.259 & 59\% \\
     & 1.5 & 0.200 & 45\% \\
\hline
4.95 & 0   & 0.465 & 81\% \\
     & 0.6 & 0.457 & 80\% \\
     & 1.5 & 0.343 & 60\% \\
\hline
\end{array}
\]
\end{table}

\begin{table}[tb]
\caption{\label{mu}
$\Gamma(M_X^{\rm max})$ with charged-lepton energy cuts
($m_b =4.8 \mbox{ GeV}$ and $\mu_\pi^2=0.6\mbox{ GeV}^2$). }
\[
\begin{array}{cccc}
\hline
M_X^{\rm max} \mbox{(GeV)} & E_l^{\rm cut} \mbox{(GeV)} & \Gamma(M_X^{\rm max})
        (|V_{ub}|^2 \mbox{ GeV}) & \mbox{percentage} \\
\hline
{\rm no~cut}  & 0   & 0.519 & 100\% \\
              & 0.6 & 0.505 & 97\% \\
              & 1.5 & 0.351 & 68\% \\
\hline
1.86 & 0   & 0.460 & 89\% \\
     & 0.6 & 0.450 & 87\% \\
     & 1.5 & 0.329 & 63\% \\
\hline
1.5  & 0   & 0.391 & 75\% \\
     & 0.6 & 0.384 & 74\% \\
     & 1.5 & 0.290 & 56\% \\
\hline
\end{array}
\]
\end{table}

\begin{table}[tb]
\caption{\label{ssah3}
$\Gamma(M_X^{\rm max})$ with charged-lepton energy cuts
($m_b =4.8 \mbox{ GeV}$ and $\alpha_s=0.35$). }
\[
\begin{array}{cccc}
\hline
M_X^{\rm max} \mbox{(GeV)} & E_l^{\rm cut} \mbox{(GeV)} & \Gamma(M_X^{\rm max})
        (|V_{ub}|^2 \mbox{ GeV}) & \mbox{percentage} \\
\hline
{\rm no~cut}   & 0   & 0.441 & 100\% \\
               & 0.6 & 0.429 & 97\% \\
              & 1.5 & 0.292 & 66\% \\
\hline
1.86 & 0   & 0.367 & 83\% \\
     & 0.6 & 0.360 & 82\% \\
     & 1.5 & 0.264 & 60\% \\
\hline
1.5  & 0   & 0.272 & 62\% \\
     & 0.6 & 0.269 & 61\% \\
     & 1.5 & 0.211 & 48\% \\
\hline
\end{array}
\]
\end{table}

\begin{table}[tb]
\caption{\label{hah1} $\Gamma_{\rm total}$ with charged-lepton energy cuts
(ACCMM distribution function of Eq. (21)).} \[
\begin{array}{cccc}
\hline
m_b \mbox{(GeV)} & E_l^{\rm cut} \mbox{(GeV)} & \Gamma_{\rm total}
         (|V_{ub}|^2 \mbox{ GeV}) & \mbox{percentage} \\
\hline
4.8  & 0   & 0.503 & 100\% \\
     & 0.6 & 0.489 & 97\% \\
     & 1.5 & 0.336 & 67\% \\
     & 2.3 & 0.055 & 11\% \\
\hline
4.65 & 0   & 0.443 & 100\% \\
     & 0.6 & 0.429 & 97\% \\
     & 1.5 & 0.287 & 65\% \\
     & 2.3 & 0.040 &  9\% \\
\hline
4.95 & 0   & 0.573 & 100\% \\
     & 0.6 & 0.558 & 97\% \\
     & 1.5 & 0.394 & 69\% \\
     & 2.3 & 0.079 & 14\% \\
\hline
\end{array}
\]
\end{table}

\begin{table}[tb]
\caption{\label{hah3}
$\Gamma(M_X^{\rm max}=1.50 \mbox{GeV})$ with charged-lepton energy cuts
(ACCMM). } \[
\begin{array}{cccc} 
\hline
m_b \mbox{(GeV)} & E_l^{cut} \mbox{(GeV)} & \Gamma(m_X^{max}=1.50
        \mbox{ GeV}) (|V_{ub}|^2 \mbox{ GeV}) & \mbox{percentage} \\
\hline
4.8  & 0   & 0.351 & 70\% \\
     & 0.6 & 0.345 & 69\% \\
     & 1.5 & 0.263 & 52\% \\
\hline
4.65 & 0   & 0.271 & 61\% \\
     & 0.6 & 0.267 & 60\% \\
     & 1.5 & 0.205 & 46\% \\
\hline
4.95 & 0   & 0.465 & 81\% \\
     & 0.6 & 0.456 & 80\% \\
     & 1.5 & 0.343 & 60\% \\
\hline
\end{array}
\]
\end{table}

%
%
\newpage
\begin{figure}

\begin{center}
\caption{ The
3-dimensional plot of 
the double differential decay rate $d\Gamma/dE_l dM_X$
with $m_b = 4.8$~GeV.
The lines in the bottom plane show the contour lines.
The three straight lines in contour figure, which are the projections of
three curved lines in the 3-dimensional figure, indicate 
the charged-lepton energy cuts,
$i.e.$ $E_l^{\rm cut} = 0.6, 1.5, 2.3$ GeV, respectively.
}
\end{center}

\begin{center}
\vspace{1cm}
\caption{
Hadronic invariant mass distribution
with $m_b=$ 4.8, 4.65, 4.95 GeV
and $E_l^{\rm cut}=$ 0, 0.6, 1.5, 2.3 GeV.
The four curves with the same line style are correspondsing to
$E_l^{\rm cut}=$ 0, 0.6, 1.5 and 2.3 GeV, respectively, from top to bottom.}
\end{center}

\begin{center}
\vspace{1cm}
\caption{ 
$\Gamma(M_X^{\rm max})$ with $m_b=$ 4.8, 4.65, 4.95 GeV
and $E_l^{\rm cut}=$ 0, 0.6, 1.5, 2.3 GeV, integrated up to $M_X^{\rm max}$.
The four curves with the same line style are correspondsing to
$E_l^{\rm cut}=$ 0, 0.6, 1.5 and 2.3 GeV, respectively, from top to bottom.}
\end{center}

\end{figure}

\newpage

\hspace{-5cm}\psfig{file=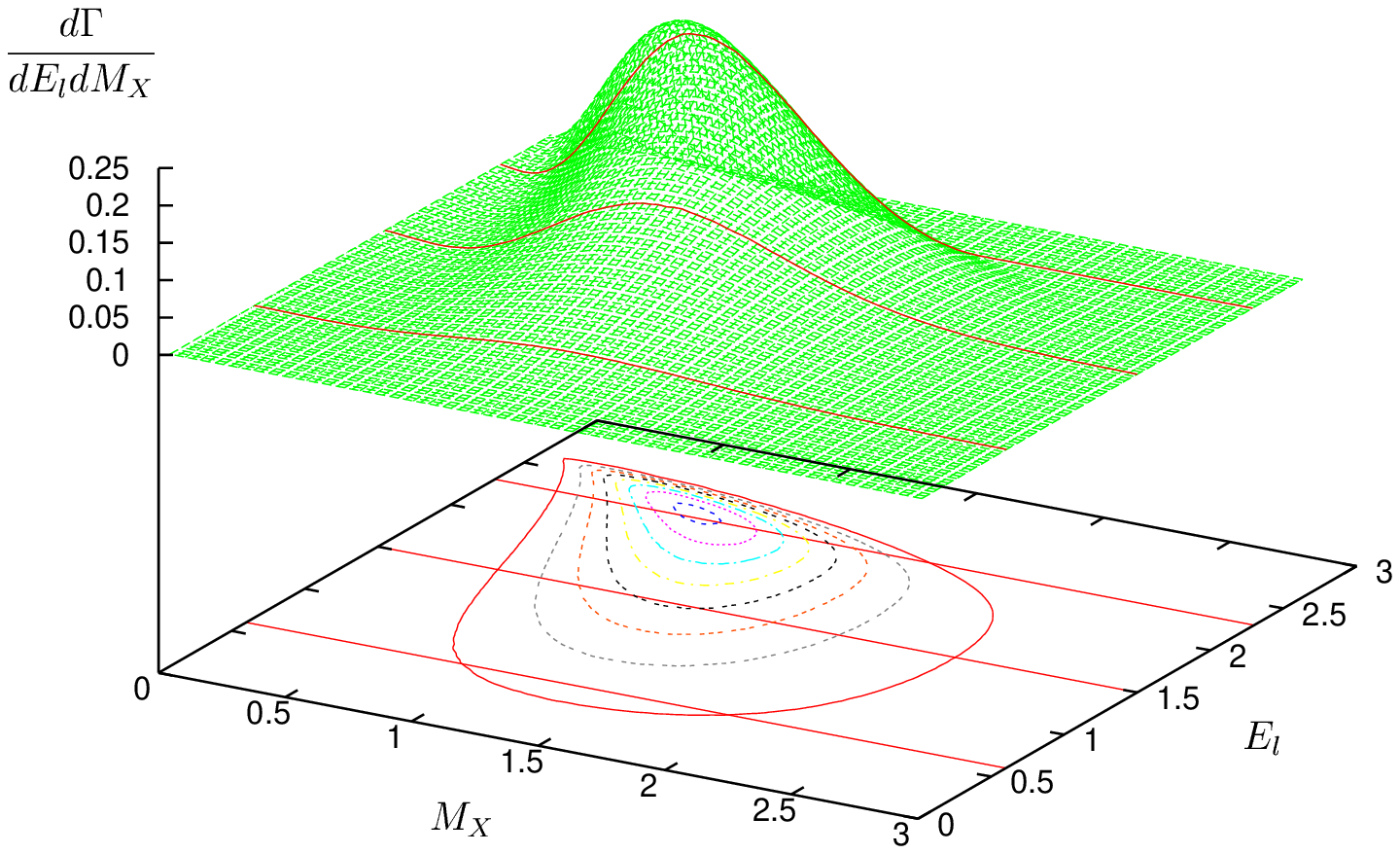}

\vspace{-5cm} FIG. 1

\newpage

\hspace{-5cm}\psfig{file=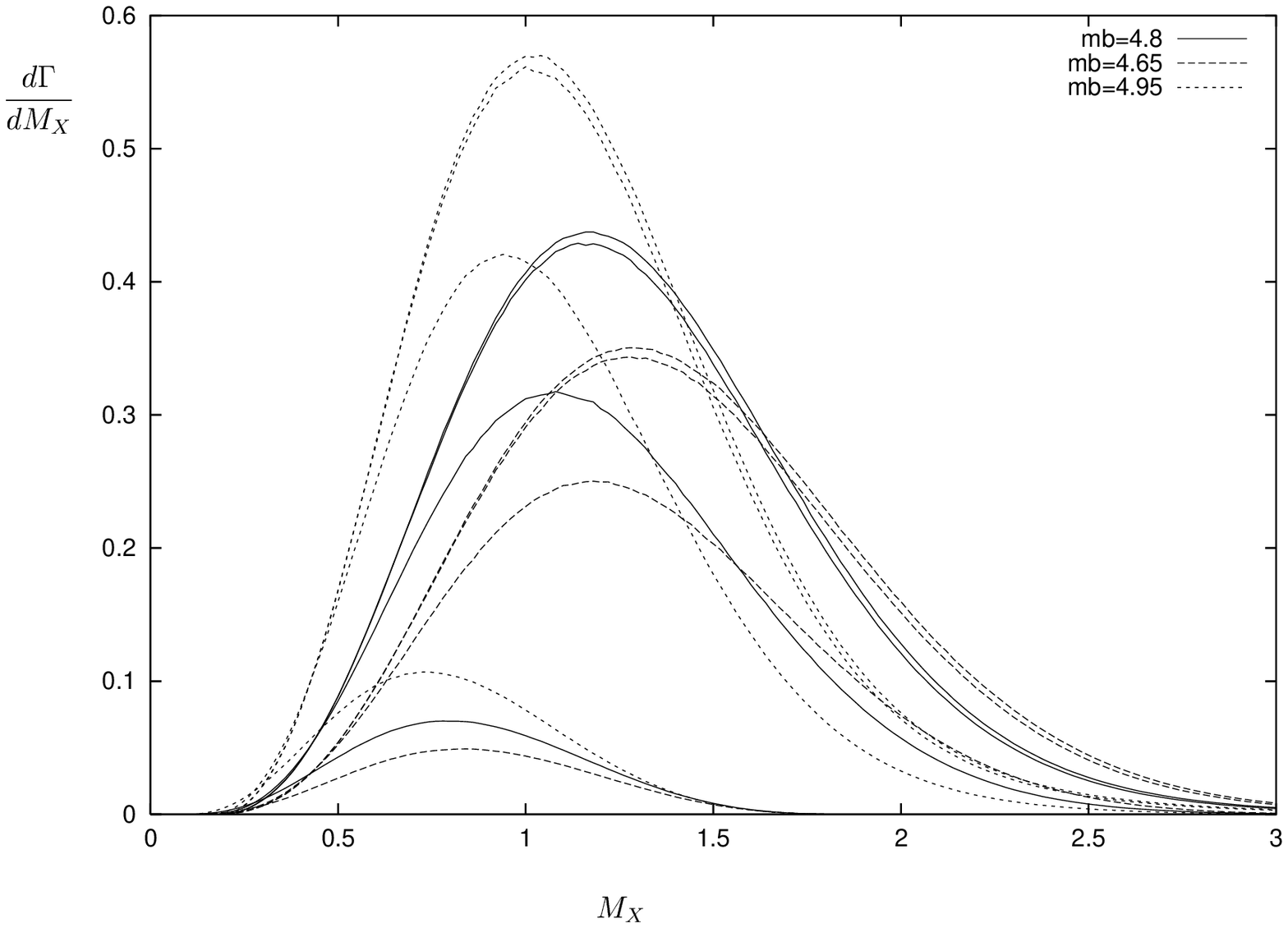}

\vspace{-5cm} FIG. 2

\newpage

\hspace{-5cm}\psfig{file=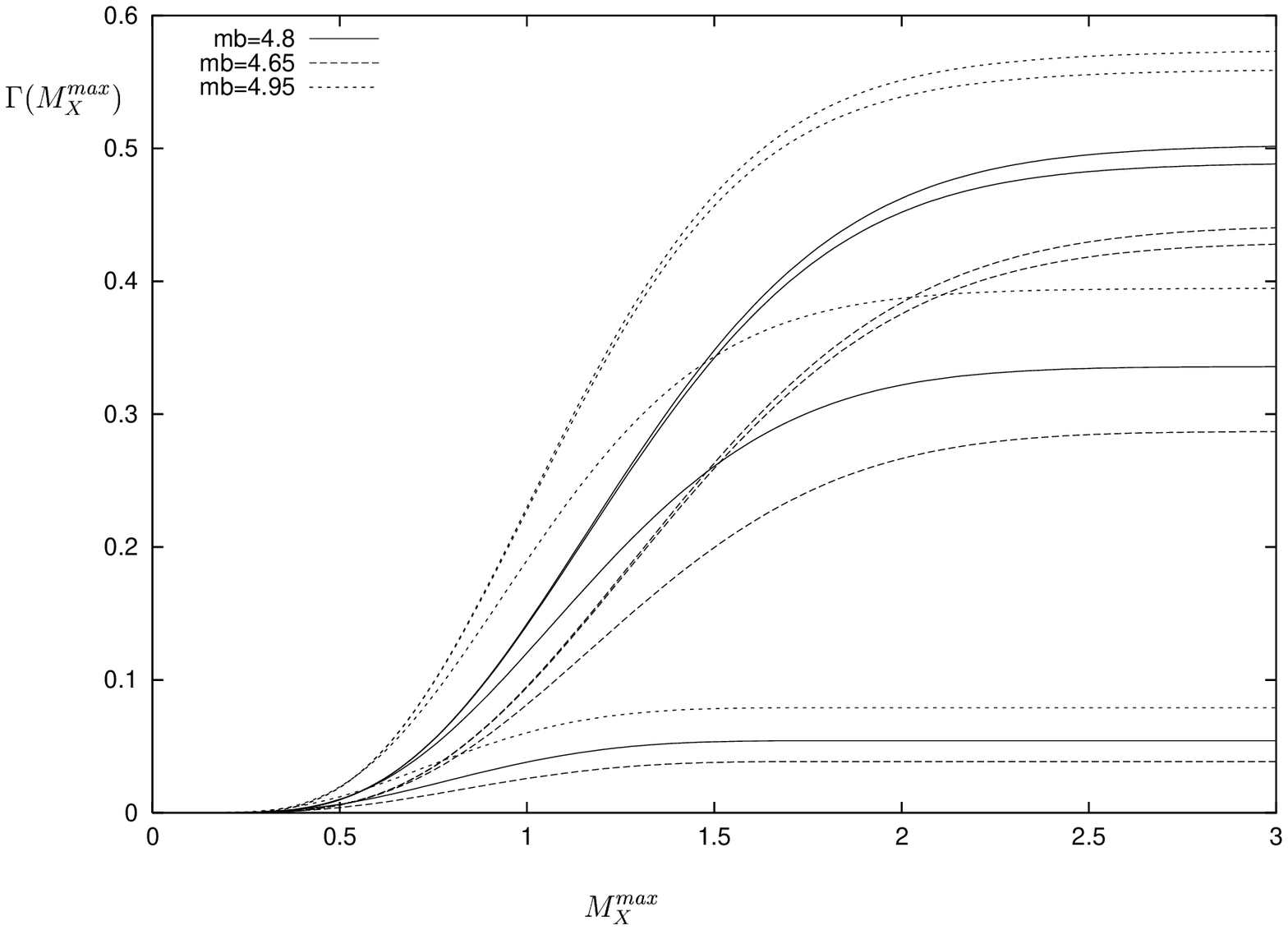}

\vspace{-5cm} FIG. 3

\end{document}